\documentstyle[eqsecnum,preprint,aps,tighten,psfig]{revtex}

\begin{document}
\draft


\title{Towards a test particle description of transport processes for states with 
continuous mass spectra}

\author{Stefan Leupold}

\address{Institut f\"ur Theoretische Physik, Justus-Liebig-Universit\"at
Giessen,\\
D-35392 Giessen, Germany}

\date{September 30, 1999}

\maketitle

\begin{abstract}
Based on a first order gradient expansion a consistent transport equation is derived for
a nonrelativistic system beyond the quasiparticle approximation, i.e.~for a regime where
the dynamically generated width of the states is allowed to be large. An exactly 
conserved quantity is identified which is interpreted as an effective particle 
number obtained by coarse graining. Using a test particle ansatz for this conserved 
quantity allows to rewrite the transport equation
into equations of motion for test particles. The two-body collision terms are formulated
in terms of the test particles which gain non-trivial renormalization factors due to the 
coarse graining process.
\end{abstract}

\section{Introduction} \label{sec:intro}

Since the pioneering works on non-equilibrium quantum field theory 
\cite{Sc61,KB,BM63,Ke64} semiclassical transport theory has become a major tool to 
solve various problems of many-body physics. While former works have focused their 
attention more or less on the quasi-particle regime 
(see e.g.~\cite{Da84a,Ch85,BotMal,DaM,MrHe,Mr,GrLe98} and references therein) and small 
corrections to it \cite{SL95,Kra,Reh} the extension of
the formalism to off-shell phenomena has become a topic of growing interest in the last
few years. E.g.~in heavy-ion collisions it has turned out that the collision rates 
usually are so high that an on-shell approximation seems to be inappropriate 
\cite{henning,IKV99,CJ99,EM99}. In addition, the resonances excited during the
reaction may have large decay width. Therefore, a representation of these states by
stable particles may not be a proper approximation. 

In this article we shall set up a
formalism which allows for a consistent transport theoretical description of
states with a continuous mass spectrum. For simplicity we restrict ourselves to the
non-relativistic case. We will comment in the last section on possible generalizations
to relativistic systems. We will derive a transport equation in first
order gradient expansion which includes off-shell effects. We will discuss in detail
how this transport equation can be solved by a proper test particle ansatz. To account 
properly for the off-shell behavior the test particles are allowed to have an arbitrary 
energy not restricted by the mass-shell condition. While a test
particle representation of a transport equation in the quasi-particle limit is 
straightforward we will see that this becomes a rather non-trivial issue if off-shell
effects are included. Especially we have to worry about a quantity which is fully
conserved by the transport equation. Only such a quantity can be represented by 
test particles. Otherwise the test particle ansatz would yield a number of equations 
which is higher than the number of test particle coordinates, i.e.~an overdetermined 
system of equations. We will isolate a quantity which is fully conserved by the
transport equation and derive the corresponding test particle equations of motion.
Finally we will comment on the form of the collision terms for these test particles.

The article is organized as follows. In the next section we discuss in some detail 
for a simple example how a test particle ansatz works for a conserved quantity and fails
for a non-conserved one. In Sec.~\ref{sec:KBge} equations of motion for the
quantities of interest are derived from the underlying quantum field theoretical
Kadanoff-Baym equations by performing a first order gradient expansion. This derivation
is very similar to the one presented recently in \cite{IKV99} (second reference). 
Therefore, Sec.~\ref{sec:KBge} might be seen as a review. It is included to keep the 
article self-contained and to put emphasis on various other aspects as compared to 
previous works. In Sec.~\ref{sec:ppn} new material is presented. Especially a
quantity is isolated which is conserved by the transport equation derived in the 
preceeding section. For this quantity a test particle ansatz is presented in 
Sec.~\ref{sec:tpeom} and the
equations of motion for the test particles are derived. Finally we evaluate the
self-energies in the Born approximation using the test particle representation. 
We will summarize our results in Sec.~\ref{sec:sum}, compare them to previous
works and give an outlook on unresolved problems.


\section{Test particle ansatz} \label{sec:simple}

Before turning to the proper transport equation for modes with a continuous mass spectrum
we would like to discuss how a test particle ansatz for the
quantity of interest works and for which cases it makes sense. For this aim let us 
consider an equation for a quantity 
$f(t,p)$ which is too complicated to solve it exactly. The only thing we have
to know about this exact equation is that it conserves a particle number
\begin{equation}
  \label{eq:simp1}
N(t) = \int\!\! dp \, f(t,p)   \,.
\end{equation}
By an approximation scheme one obtains from the exact equation the following approximate
one (which resembles a Vlasov equation):
\begin{equation}
  \label{eq:simp2}
{\partial \over \partial t} \left[ ( 1 - \kappa(t,p)) f(t,p) \right] + 
F(t) \, {\partial \over \partial p} f(t,p) = 0   
\end{equation}
with a force term $F(t)$ and a renormalization $\kappa$ which both are not further 
specified. Obviously this approximate (transport) equation does {\it not} conserve
the particle number $N$ but instead the quantity 
\begin{equation}
  \label{eq:simp3}
\tilde N(t) = \int dp \, (1-\kappa(t,p)) f(t,p)   \,.
\end{equation}
We now discuss two cases
to solve (\ref{eq:simp2}),
namely test particle ans\"atze for $f$ and for $\tilde f = (1-\kappa)f$, respectively. 
In general, a test particle representation for any of the two quantities is given by
\begin{equation}
  \label{eq:simp4}
\left.
  \begin{array}{c}
f \\ \mbox{or} \\ \tilde f
  \end{array}
\right\} =
{1 \over L} \sum\limits_{i=1}^M \delta(p-p_i(t))
\end{equation}
where $L$ is a normalization constant, $i$ numbers the $M$ test particles and $p_i(t)$ 
characterizes the trajectory of the test particle $i$. Obviously the momentum integral
over (\ref{eq:simp4}) is conserved and given by $M/L$. Therefore, a test particle ansatz
for $f$ to solve (\ref{eq:simp2}) seems to be inappropriate since the momentum integral
over $f$ (which yields the particle number) is not conserved by (\ref{eq:simp2}). 
Nonetheless, it is  instructive to figure out which equations one gets with a test
particle ansatz for $f$. In this case one finds
\begin{equation}
0 = \sum\limits_i  \left\{ - {d \kappa(t,p_i(t)) \over dt} \, \delta(p-p_i(t))
+ \left[\vphantom{\int}
F(t)-[1-\kappa(t,p_i(t))]\dot p_i(t)\right] {\partial \over \partial p} \delta(p-p_i(t))
\right\}  \,.   \label{eq:simp5}
\end{equation}
Note the appearance of the total derivative with respect to $t$ in contrast to the 
partial derivative in (\ref{eq:simp2}). This is obtained by
\begin{eqnarray}
\lefteqn{{ \partial \over \partial t} [(1-\kappa(t,p))\,\delta(p-p_i)] =
-{ \partial \kappa(t,p) \over \partial t} \, \delta(p-p_i)
+ (1-\kappa(t,p)) \,{ \partial \over \partial t} \delta(p-p_i) } \nonumber \\ &&
= -{ \partial \kappa(t,p) \over \partial t} \, \delta(p-p_i)
+ (1-\kappa(t,p)) \,{ \partial \over \partial p} \delta(p-p_i) \, (-\dot p_i) 
\nonumber \\ &&
= -{ \partial \kappa(t,p_i) \over \partial t} \, \delta(p-p_i) 
- \dot p_i \,{ \partial \kappa(t,p_i) \over \partial p_i} \, \delta(p-p_i) 
- (1-\kappa(t,p_i))\, \dot p_i \, { \partial \over \partial p} \delta(p-p_i)
\nonumber \\ &&
= -{ d \kappa(t,p_i) \over dt} \, \delta(p-p_i) 
- (1-\kappa(t,p_i))\, \dot p_i \, { \partial \over \partial p} \delta(p-p_i)
  \label{eq:long}
\end{eqnarray}
where we have used the identity
\begin{equation}
  \label{eq:ident}
h(x) \, {\partial \over \partial x} \delta(x-y) 
= h(y) \, {\partial \over \partial x} \delta(x-y) - {d h(y) \over dy} \, \delta(x-y)  \,.
\end{equation}
To fulfill (\ref{eq:simp5}) one has to demand that the coefficient of the 
$\delta$-function as well as the coefficient of the momentum derivative of the 
$\delta$-function both vanish, i.e.
\begin{eqnarray}
  \label{eq:two1}
- {d \kappa(t,p_i(t)) \over dt} &=& 0  \,, \\
F(t)-[1-\kappa(t,p_i(t))]\dot p_i(t) &=& 0  \,.
\end{eqnarray}
In this way we obtain {\it two} equations of motion for {\it one} test particle 
coordinate:
\begin{eqnarray}
  \label{eq:two2}
\dot p_i(t) &=& 
-{ {\displaystyle \partial \kappa(t,p_i) \over \displaystyle \partial t} \over 
{\displaystyle \partial \kappa(t,p_i) \over \displaystyle \partial p_i} } \,, \\
  \label{eq:two3}
\dot p_i(t) &=& { F(t) \over 1 - \kappa(t,p_i)}   \,.
\end{eqnarray}
Obviously, this provides an overdetermined system which is simply due to the fact that
a test particle ansatz has been made for a quantity which is not conserved by the 
equation one wants to solve. On the other hand, a test particle ansatz for $\tilde f$
yields
\begin{eqnarray}
0 &=& \sum\limits_i  \left[\vphantom{\int}
{F(t) \over 1 - \kappa(t,p_i(t)) } - \dot p_i(t) \right] 
{\partial \over \partial p} \delta(p-p_i(t))
\end{eqnarray}
Here the coefficient of the pure $\delta$-function vanishes. This is a general property
if a test particle ansatz for a conserved quantity is made. We only have to fulfill one
equation:
\begin{equation}
  \label{eq:tpftild}
\dot p_i(t) = { F(t) \over 1 - \kappa(t,p_i)}   \,.
\end{equation}
Here we observe a second interesting feature of a test particle ansatz. The last 
equation is identical to the second one of the equations obtained before. Both 
equations (\ref{eq:two3}) and (\ref{eq:tpftild}) were obtained by
demanding that the respective coefficient of the $\partial_p \delta(p-p_i)$ term has to 
vanish. Thus, the difference between a test particle ansatz for a conserved quantity
and for a non-conserved quantity lies only in the appearance of a term with a pure 
$\delta$-function (i.e.~without derivatives acting on it). This provides an additional
equation which causes the system of equations to be overdetermined for the case of a 
non-conserved quantity. The other equations do not differ for both cases. 

So far we were only concerned with a proper solution of the transport equation 
(\ref{eq:simp2}) and indeed we have presented one consistent way to solve it, namely by 
a test particle ansatz for $\tilde f$. 
Let us now recall that (\ref{eq:simp2}) is only an approximate equation and that the 
exact equation
in contrast to the transport equation exactly conserves $f$. Thus, one has to
realize that the two equations (\ref{eq:two2},\ref{eq:two3}) are identical to each 
other up to terms which
have been neglected when the transport equation (\ref{eq:simp2}) was derived from the 
underlying exact
equation. Therefore, one might conclude that one can use one or the other equation to
evolve the test particle coordinates and thus $f$ in time. The difference between these
two evolutions is in an order which was neglected anyway when deriving the transport 
equation from the exact one. Therefore, we now choose e.g.~(\ref{eq:two3}) 
to evolve $f$. One has to realize, however, that now it is no longer
the transport equation (\ref{eq:simp2}) one is solving. Instead, one can reconstruct 
from the test particle equation of motion that the corresponding transport equation is 
given by
\begin{eqnarray}
0 &=& \sum\limits_i  \left[\vphantom{\int}
{F(t) \over 1 - \kappa(t,p_i(t)) } - \dot p_i(t) \right] 
{\partial \over \partial p} \delta(p-p_i(t))
\nonumber \\
&=& {\partial \over \partial p} \left[ {F(t) \over 1 - \kappa(t,p) } \sum\limits_i  
\delta(p-p_i(t)) \right]
+ {\partial \over \partial t} \sum\limits_i  \delta(p-p_i(t))
\nonumber \\
  \label{eq:reconstr}
&=& F(t)\,{\partial \over \partial p} {f(t,p) \over 1 - \kappa(t,p) } 
+ {\partial \over \partial t} f(t,p) 
\end{eqnarray}
which clearly differs from (\ref{eq:simp2}). However, by construction this new 
transport equation (\ref{eq:reconstr}) is as close
to the underlying exact equation as the original transport equation (\ref{eq:simp2}). 

To summarize we have presented two ways to deal with the approximate transport equation 
and the fact that the particle number is conserved by the exact equation. 
\begin{enumerate}
\item Extract a quantity which is exactly conserved by the approximate equation 
(here $\tilde N$). Make a test particle ansatz for the corresponding density (here 
$\tilde f$). Once the transport equation is solved one can also calculate $f$ and
$N$ and figure out to which extent the conservation of $N$ is violated in time. This
provides a check for the accuracy of the approximation scheme which has served to
derive the transport equation from the exact one. 
\item Make a test particle ansatz for the quantity which is exactly conserved by the 
exact equation but not by the transport equation (here $f$). Ignore one of the obtained 
equations of motion for the test particle coordinates. Usually one is tempted to ignore
the equation obtained from the coefficient of the pure $\delta$-function. Reconstruct
a new transport equation from the test particle equations of motion. 
\end{enumerate}

At first sight the second approach seems to be more appealing since there the 
reconstructed transport equation shares with the exact equation the full conservation 
of the particle number. Unfortunately things are not always as simple as the chosen 
example seems to indicate. The problem is that in general it is not always possible to 
reconstruct such a transport equation from the test particle equations. As long as only
the propagation of test particles is concerned (Vlasov-type equation) one can 
reconstruct a corresponding transport equation as shown above. If, however, also 
collisions of test particles are taken into account (Boltzmann-type equations) new
problems arise. To see this we extend (\ref{eq:simp2}) by adding a collision term on 
the r.h.s.
\begin{equation}
  \label{eq:simp2wcoll}
{\partial \over \partial t} \left[ ( 1 - \kappa(t,p)) f(t,p) \right] + 
F(t) \, {\partial \over \partial p} f(t,p) = I_{\rm coll}[f]  
\end{equation}
with the additional condition 
\begin{equation}
  \label{eq:condcoll}
\int\!\! dp \, I_{\rm coll}[f] = 0   \,.
\end{equation}
We stress again that this equation has to be regarded as an approximate (transport) 
equation. For the underlying exact equation we assume that it is too complicated to solve
and that it conserves the particle number $N$. 
Again we observe that $\tilde N$ is conserved by (\ref{eq:simp2wcoll}) while $N$ is not 
conserved. 
The most pragmatic (albeit doubtful) approach would be the following: 
One makes a test particle ansatz
for $f$. For the propagation of the test particles {\it between collisions} one obtains
again the overdetermined system of equations (\ref{eq:two2},\ref{eq:two3}). 
One disregards the first equation and 
uses the second one to evolve the test particle coordinates in time. One allows in 
addition for collisions between the particles according to the collision term on the
r.h.s.~of (\ref{eq:simp2wcoll}). The obstacle in that pragmatic approach is the fact 
that there is {\it no}
corresponding transport equation at all which can be derived from (\ref{eq:simp2wcoll}).
Reconstructing a transport equation from the test particle equation of motion yields
\begin{equation}
  \label{eq:reconstwcoll}
{\partial \over \partial t} f(t,p) + 
F(t)\,{\partial \over \partial p} {f(t,p) \over 1 - \kappa(t,p) } 
= {I_{\rm coll}[f]  \over 1 - \kappa(t,p) } \,.
\end{equation}
Thus the collision term now is given by $I_{\rm coll}/(1-\kappa)$ instead of 
$I_{\rm coll}$. This new collision term, however, does not conserve the particle number 
\begin{equation}
  \label{eq:nocpwcoll}
{d N \over dt} = {d \over dt} \int\!\! dp \, f(t,p) 
= \int\!\! dp \, {I_{\rm coll}[f]  \over 1 - \kappa(t,p) } \neq 0  \,.
\end{equation}
Therefore, the reconstructed transport equation (\ref{eq:reconstwcoll}) has no 
advantages as compared to the original one (\ref{eq:simp2wcoll}). Using instead the 
original collision term $I_{\rm coll}$ as described above 
is only a guess. It is not clear whether such an ad hoc postulated equation is as close
to the exact equation as the transport equation (\ref{eq:simp2wcoll}). 

Aiming at a field theoretical foundation for the description of transport processes we 
think it is not acceptable to have only a recipe for the test particle evolution which is
not derivable from the underlying theory. If an
already approximate transport equation is only approximately solved one loses the
contact to the original exact equation more and more in a not controllable way. 
Therefore, we suggest to use the first approach described above, i.e.~the test particle
ansatz for $\tilde f$, since it solves the
transport equation exactly. In addition, it provides the possibility to judge the 
approximation scheme which has led to the transport equation simply by evaluating the 
time dependence of the quantity which is conserved by the exact equation.

After these precursory considerations concerning the use of the test particle ansatz
we now turn to the derivation of a transport equation from the underlying 
field-theoretical Kadanoff-Baym equations.


\section{Kadanoff-Baym equations and gradient expansion} \label{sec:KBge}

Following Kadanoff and Baym \cite{KB} we start with the exact nonrelativistic equations 
of motion for the two-point functions
\begin{eqnarray}
  \label{eq:exeom1}
\left(i {\partial \over \partial t_1} + {\Delta_1 \over 2m}\right) D^<(1,1') &=& 
\int \!\! d\bar 1 
\left[
    \Sigma^{\rm ret}(1,\bar 1) \, D^<(\bar 1, 1') 
  + \Sigma^<(1,\bar 1) \, D^{\rm av}(\bar 1, 1')
\right]  
\,, \\
  \label{eq:exeom2}
\left(i {\partial \over \partial t_1} + {\Delta_1 \over 2m}\right) D^>(1,1') &=& 
\int \!\! d\bar 1 
\left[ 
    \Sigma^{\rm ret}(1,\bar 1) \, D^>(\bar 1, 1') 
  + \Sigma^>(1,\bar 1) \, D^{\rm av}(\bar 1, 1')
\right]  
\end{eqnarray}
where we have introduced the two-point functions without ordering
\begin{eqnarray}
  \label{eq:defdless}
i D^< (x,y) &=& \pm \langle \psi^\dagger(y) \,\psi(x) \rangle  \,, \\
  \label{eq:defdmore}
i D^> (x,y) &=& \phantom{\pm} \langle \psi(x) \,\psi^\dagger (y) \rangle 
\end{eqnarray}
and the retarded and advanced quantities
\begin{eqnarray}
  \label{eq:defdret}
D^{\rm ret}(x,y) = \Theta(x_0-y_0) \left[ D^>(x,y) -D^<(x,y) \right] \,, \\
  \label{eq:defdav}
D^{\rm av}(x,y) = \Theta(y_0-x_0) \left[ D^<(x,y) -D^>(x,y) \right]  \,.
\end{eqnarray}
In (\ref{eq:defdless}) and throughout this work the upper (lower) sign refers to bosons 
(fermions) except where otherwise stated. The self-energies are connected via
\begin{eqnarray}
  \label{eq:defsigmaretav}
\Sigma^{\rm ret}(x,y) = \Sigma^{\rm HF}(x,y) +
         \Theta(x_0-y_0) \left[ \Sigma^>(x,y) -\Sigma^<(x,y) \right] \,, \\
\Sigma^{\rm av}(x,y) = \Sigma^{\rm HF}(x,y) +
         \Theta(y_0-x_0) \left[ \Sigma^<(x,y) -\Sigma^>(x,y) \right]  \,.
\end{eqnarray}
The collisional self-energies $\Sigma^<$ and $\Sigma^>$ will be specified below. 
We have also introduced the time-local Hartree-Fock 
self-energy $\Sigma^{\rm HF}(x,y) \sim \delta(x_0-y_0)$.
In (\ref{eq:exeom1},\ref{eq:exeom2}) the numbers denote short-hand notations for the
space-time coordinates. 

In the following we will concentrate on systems where the dependence of an arbitrary 
two-point function or self-energy $F(x,y)$ on its center-of-mass variable $(x+y)/2$ is 
weak (albeit not negligible). Therefore it is appropriate to introduce the Fourier 
transform with respect to the more rapidly oscillating difference variable $x-y$ (Wigner 
transformation):
\begin{equation}
  \label{eq:wigdef}
\bar F(X,p) = \int \!\! d^4\!u \, e^{ipu} F(X+u/2,X-u/2)
\end{equation}
where $F$ denotes an arbitrary two-point function or self-energy,
\begin{equation}
F = D^<,\, D^>,\dots,\, \Sigma^<,\dots  \,.
\end{equation}

From the definitions 
(\ref{eq:defdless},\ref{eq:defdmore},\ref{eq:defdret},\ref{eq:defdav},\ref{eq:wigdef})
we obtain the following transformation properties with respect to complex conjugation:
\begin{eqnarray}
\left[\bar D^<(X,p)\right]^* &=& - \bar D^<(X,p)\,,  \nonumber\\
  \label{eq:trafocc}
\left[\bar D^>(X,p)\right]^* &=& - \bar D^>(X,p) \,,  \\ 
\left[\bar D^{\rm ret}(X,p)\right]^{*} &=& \bar D^{\rm av}(X,p)  \,. \nonumber
\end{eqnarray}
Corresponding equations hold also for the self-energies. It is useful to introduce the 
real-valued quantities 
\begin{equation}
  \label{eq:defsless}
S^<(X,p) = \pm i \bar D^<(X,p) \,, \qquad S^>(X,p) = i \bar D^>(X,p)
\end{equation}
and 
\begin{eqnarray}
  \label{eq:defspec}
{\cal A}(X,p) &=& -2 {\rm Im}\bar D^{\rm ret}(X,p) = 2 {\rm Im}\bar D^{\rm av}(X,p)
 \nonumber \\ &=& i[\bar D^{\rm ret}(X,p) - \bar D^{\rm av}(X,p)]  
= i[\bar D^>(X,p) - \bar D^<(X,p)] 
\nonumber \\ &=& S^>(X,p) \mp S^<(X,p) \,.
\end{eqnarray}
The quantity ${\cal A}$ is the spectral function. On account of the (anti-)commutation
relation for bosonic (fermionic) fields 
\begin{equation}
  \label{eq:qcomrel}
\psi(t,\vec x) \,\psi^\dagger (t,\vec y) \mp \psi^\dagger(t,\vec y) \,\psi(t,\vec x)
= \delta(\vec x - \vec y) 
\end{equation}
the spectral function is normalized to one:
\begin{equation}
  \label{eq:normspec}
\int \!\! {dp_0 \over 2\pi} {\cal A}(X,p) = 1  \,.
\end{equation}
The quantity $S^<$ can be
interpreted as a generalized phase-space density. The particle number can be obtained
from $S^<$ as
\begin{equation}
  \label{eq:partnumbexact}
N(t) = \int\!\! d^3\!x \int\!\!{d^4\!p \over (2\pi)^4} S^<(t,\vec x;p)  \,.
\end{equation}
For a system which has reached thermal equilibrium $S^<(X,p)$ does no longer depend on
$X$ and is given by
\begin{equation}
  \label{eq:thermal}
S^<_{\rm th}(p) = n_{\rm B,F}(p_0) \, {\cal A}(p) 
\end{equation}
where $n_{\rm B,F}$ is the thermal Bose (Fermi) distribution. 

For a general 
off-equilibrium situation we are aiming at equations of motion for the real-valued 
quantities $S^<$, ${\cal A}$, 
and ${\rm Re}\bar D^{\rm ret} = {\rm Re}\bar D^{\rm av}$. (Note that $S^>$ can be
obtained from $S^<$ and ${\cal A}$ according to (\ref{eq:defspec})). In principle,
${\rm Re}\bar D^{\rm ret}$ can be obtained from ${\cal A}$ via a dispersion relation
which can be easily obtained by Wigner transformation of the definitions 
(\ref{eq:defdret},\ref{eq:defdav}):
\begin{equation}
  \label{eq:dispdret}
{\rm Re}\bar D^{\rm ret}(X,p) = 
\int\!\! {dk_0 \over 2 \pi} {\cal P}{1 \over p_0-k_0} {\cal A}(X;k_0,\vec p)
\end{equation}
where ${\cal P}$ denotes the principal value.
It is easier, however, to derive the equation of motion for ${\rm Re}\bar D^{\rm ret}$ 
from an equation of motion for $D^{\rm ret}$. Using (\ref{eq:exeom1},\ref{eq:exeom2}) and
the definition (\ref{eq:defdret}) we get
\begin{equation}
  \label{eq:exeomret}
\left(i {\partial \over \partial t_1} + {\Delta_1 \over 2m}\right) D^{\rm ret}(1,1') = 
\delta^{(4)}(1,1') 
+ \int \!\! d\bar 1 \, \Sigma^{\rm ret}(1,\bar 1) \, D^{\rm ret}(\bar 1,1')
\,.
\end{equation}
As we will see below the final result for ${\rm Re}\bar D^{\rm ret}$ agrees with 
(\ref{eq:dispdret}).

As already mentioned we are interested in systems where the dependence of the two-point
functions on $X$ is weak. This allows to Wigner transform the equations of motion
(\ref{eq:exeom1},\ref{eq:exeom2},\ref{eq:exeomret}) and to neglect all terms with more 
than one derivative
with respect to the center-of-mass variable $X$. We get the following equations
of motion in first order gradient expansion:
\begin{eqnarray}
  \label{eq:eomgrad1}
\lefteqn{\left(p_0 - {\vec p^2 \over 2m} \right) \bar D^<(X,p) 
- {i \over 2} \, [p_0 - {\vec p^2 \over 2m} ,\, \bar D^<(X,p) ] = } \nonumber \\
&&  \bar \Sigma^{\rm ret}(X,p) \, \bar D^<(X,p) 
+ \bar \Sigma^<(X,p) \, \bar D^{\rm av}(X,p)     \nonumber \\
&& {}- {i \over 2} \, [\bar \Sigma^{\rm ret}(X,p) ,\, \bar D^<(X,p) ] 
- {i \over 2} \, [\bar \Sigma^<(X,p), \, \bar D^{\rm av}(X,p) ]    \,,
\\  \nonumber \\
  \label{eq:eomgrad2}
\lefteqn{\left(p_0 - {\vec p^2 \over 2m} \right) \bar D^>(X,p) 
- {i \over 2} \, [p_0 - {\vec p^2 \over 2m} ,\, \bar D^>(X,p) ] = } \nonumber \\
&&  \bar \Sigma^{\rm ret}(X,p) \, \bar D^>(X,p) 
+ \bar \Sigma^>(X,p) \, \bar D^{\rm av}(X,p)     \nonumber \\
&& {}- {i \over 2} \, [\bar \Sigma^{\rm ret}(X,p) ,\, \bar D^>(X,p) ] 
- {i \over 2} \, [\bar \Sigma^>(X,p), \, \bar D^{\rm av}(X,p) ]    
\\ \nonumber \\
  \label{eq:eomgradret}
\lefteqn{\left(p_0 - {\vec p^2 \over 2m} \right) \bar D^{\rm ret}(X,p) 
- {i \over 2} \, [p_0 - {\vec p^2 \over 2m} ,\, \bar D^{\rm ret}(X,p) ] = } \nonumber \\
&& 1 + \bar \Sigma^{\rm ret}(X,p) \, \bar D^{\rm ret}(X,p)
- {i \over 2} \, [\bar \Sigma^{\rm ret}(X,p) ,\, \bar D^{\rm ret}(X,p) ] 
\end{eqnarray}
where we have introduced the generalized Poisson bracket 
\begin{equation}
  \label{eq:poisson}
[A,B] = \partial_{X_0} A \,\partial_{p_0} B - \partial_{p_0} \,A \partial_{X_0} B 
- \vec\nabla_X A \,\vec\nabla_p B + \vec\nabla_p A \,\vec\nabla_X B   \,.
\end{equation}
It is worth noting that the drift terms typical for transport equations show up in the
expressions
\begin{equation}
  \label{eq:driftstand}
[p_0 - {\vec p^2 \over 2m} ,\, B] = -\partial_{X_0} B - {\vec p \over m}
\cdot \vec \nabla_X B   \,.
\end{equation}
We note that for the derivation of (\ref{eq:eomgrad1}-\ref{eq:eomgradret}) we have not 
assumed anything about the derivatives with respect to $p$. This is different
to the approach of \cite{BotMal} where systems were studied for which
mixed double derivatives like $\vec\nabla_X \vec\nabla_p A$ are also negligibly small.
We will come back to that subtle difference below.

Next we extract from (\ref{eq:eomgrad1}-\ref{eq:eomgradret}) real and imaginary
parts using (\ref{eq:defsless},\ref{eq:defspec}) and the decompositions
\begin{eqnarray}
  \label{eq:decsigretav}
\bar \Sigma^{\rm ret/av}(X,p) = 
{\rm Re}\bar \Sigma^{\rm ret/av}(X,p) \mp {1 \over 2}\,i\,\Gamma(X,p) 
\end{eqnarray}
where we have introduced the (real-valued, non-negative) width
\begin{equation}
  \label{eq:defgam}
\Gamma = i(\bar\Sigma^>- \bar\Sigma^<) = 
i(\bar \Sigma^{\rm ret}-\bar \Sigma^{\rm av})    \,.
\end{equation}
In (\ref{eq:decsigretav}) the minus (plus) sign refers to the retarded (advanced)
self-energy. We find
\begin{eqnarray}
  \label{eq:anasless}
\left(p_0 - {\vec p^2 \over 2m} \right) S^< &=&
{\rm Re}\bar \Sigma^{\rm ret}\,S^< \pm i \bar\Sigma^< \,{\rm Re}\bar D^{\rm ret}
-{1 \over 4}[\Gamma,\, S^<] + {1 \over 4}[\pm i\bar\Sigma^<,\,{\cal A}]    \,, \\
  \label{eq:gradsless}
[p_0 - {\vec p^2 \over 2m},\, S^<] &=& \Gamma \, S^< \mp i \bar\Sigma^< \,{\cal A}
+[{\rm Re}\bar \Sigma^{\rm ret},\, S^<] 
+ [\pm i \bar\Sigma^<, \, {\rm Re}\bar D^{\rm ret}]     \,,       \\
  \label{eq:anaspec}
\left(p_0 - {\vec p^2 \over 2m} \right) {\cal A} &=& 
{\rm Re}\bar \Sigma^{\rm ret} \, {\cal A} + \Gamma \, {\rm Re}\bar D^{\rm ret} \,, \\
  \label{eq:gradspec}
[p_0 - {\vec p^2 \over 2m},\, {\cal A}] &=& 
[{\rm Re}\bar \Sigma^{\rm ret}, \, {\cal A}] + [\Gamma, \, {\rm Re}\bar D^{\rm ret}] 
\,, \\
  \label{eq:anare}
\left(p_0 - {\vec p^2 \over 2m} \right) {\rm Re}\bar D^{\rm ret} &=& 
1+{\rm Re}\bar \Sigma^{\rm ret} \, {\rm Re}\bar D^{\rm ret} 
-{1 \over 4} \Gamma \, {\cal A}  \,,  \\
  \label{eq:gradre}
[p_0 - {\vec p^2 \over 2m},\, {\rm Re}\bar D^{\rm ret}] &=& 
[{\rm Re}\bar \Sigma^{\rm ret}, \, {\rm Re}\bar D^{\rm ret}] 
-{1 \over 4} [\Gamma, \, {\cal A}]    \,.
\end{eqnarray}
Note that $i \bar\Sigma^<$ is real-valued (cf.~Eq.~(\ref{eq:trafocc}) and the remark 
concerning self-energies).

Obviously using the gradient expansion up to first order we have obtained six equations
for three quantities ($S^<$, ${\cal A}$, and ${\rm Re}\bar D^{\rm ret}$). Therefore
some of these six equations are either redundant or provide constraints which have to
be fulfilled by the dynamical quantities to make sure that their evolution is in 
accordance with the first order gradient expansion. To say it in other words: Even if
one starts out with a configuration (represented by the two-point functions at initial
time) for which the dependence on $X$ is weak, a peculiar choice for the self-energies
might drive the system out of the regime where the gradient expansion is valid. Thus
it would not be surprising if the gradient expanded equations of motion provided 
constraints on the dynamical quantities. 

As we will see now there are indeed 
redundant equations as well as constraints. Let us start with the purely algebraic
equations (\ref{eq:anaspec},\ref{eq:anare}). They already yield the following
expressions for two of the three quantities of interest:
\begin{eqnarray}
  \label{eq:solspec}
{\cal A}(X,p) &=& { \Gamma(X,p) \over 
\left( 
p_0 - {\vec p^2 \over 2m}-{\rm Re}\bar \Sigma^{\rm ret}(X,p)
\right)^2
+ {1 \over 4} \Gamma^2(X,p)  }        \,,    \\
  \label{eq:solred}
{\rm Re}\bar D^{\rm ret}(X,p) &=& 
{ p_0 - {\vec p^2 \over 2m}-{\rm Re}\bar \Sigma^{\rm ret}(X,p)  \over 
\left( 
p_0 - {\vec p^2 \over 2m}-{\rm Re}\bar \Sigma^{\rm ret}(X,p)
\right)^2
+ {1 \over 4} \Gamma^2(X,p)  }    \,.
\end{eqnarray}
For the retarded and advanced two-point functions this yields
\begin{equation}
  \label{eq:soldretav}
\bar D^{\rm ret/av}(X,p) 
= {1 \over p_0 - {\vec p^2 \over 2m}-\bar \Sigma^{\rm ret/av}(X,p)}    \,.
\end{equation}
These relations are very well known for systems which are completely
homogeneous in space and time, i.e.~where the zeroth order in the gradient expansion is
sufficient. It is interesting to observe now that these relations are still valid in 
first order gradient expansion \cite{henning,GrLe98}. We also note that inserting 
(\ref{eq:solspec}) in (\ref{eq:dispdret}) yields (\ref{eq:solred}) provided that
the used self-energy is indeed retarded, i.e.~satisfies the dispersion relation
\begin{equation}
  \label{eq:dispsigret}
{\rm Re}\bar \Sigma^{\rm ret}(X;p_0,\vec p) = \bar \Sigma^{\rm HF}(X;\vec p) +
\int\!\! {dk_0 \over 2 \pi} {\cal P}{1 \over p_0-k_0} \Gamma(X;k_0,\vec p)   \,.
\end{equation}
Finally one can show that the expression for the spectral function (\ref{eq:solspec})
indeed fulfills the normalization condition (\ref{eq:normspec}). This is discussed
in Appendix \ref{app:psspec}.

Next, we observe that the expressions (\ref{eq:solspec},\ref{eq:solred}) identically
solve (\ref{eq:gradspec},\ref{eq:gradre}). To see this one might simply insert 
(\ref{eq:solspec},\ref{eq:solred}) and perform the somewhat tedious brute-force 
calculation. Instead we will take here the following elegant way \cite{kainu}: 
We combine (\ref{eq:gradspec},\ref{eq:gradre}) to a complex equation. 
(\ref{eq:gradre})$-{i \over 2}$(\ref{eq:gradspec}) yields
\begin{equation}
  \label{eq:trivial}
[p_0 - {\vec p^2 \over 2m}-\bar \Sigma^{\rm ret} , \, \bar D^{\rm ret}] = 0  \,.
\end{equation}
It is trivial to see that the expression (\ref{eq:soldretav}) solves this equation.
It is interesting to note here that the same reasoning holds for relativistic
scalar bosons. For relativistic fermions, however, things might be more complicated
since in this case the generalized Poisson bracket of the matrix valued quantities 
$\bar D^{\rm ret}$ and $(\bar D^{\rm ret})^{-1}$ may not vanish any more \cite{kainu}. 
To the best of our knowledge this subtlety has never been worked out in detail
(concerning a closely related case cf.~\cite{kainu2}).

We conclude that the equations (\ref{eq:gradspec},\ref{eq:gradre}) provide no
additional information as compared to (\ref{eq:anaspec},\ref{eq:anare}); they are 
redundant. Therefore, the set of equations
(\ref{eq:anasless}-\ref{eq:gradre}) contains at most four independent equations
for the three quantities of interest. 

Now we turn our attention to (\ref{eq:anasless},\ref{eq:gradsless}). Before analyzing
them for the general case of arbitrary width we briefly discuss
their meaning for the quasi-particle regime: In this limit the gradient terms in
(\ref{eq:anasless}) and also the second term on the r.h.s.~vanish and this equation 
becomes a purely algebraic (constraint) equation demanding that $S^<$ ought to be 
proportional to an on-shell $\delta$-function. In other words this equation reduces
to the mass-shell constraint discussed e.g.~in \cite{DaM,MrHe,Mr,GrLe98}.
The traditional on-shell transport equation is recovered from (\ref{eq:gradsless})
by integrating over the energy \cite{KB}.
As already pointed out the l.h.s.~provides the drift terms. The first two terms on the
r.h.s.~yield the collision terms as will be discussed in more detail below. The third term
on the r.h.s.~gives the Vlasov contribution. The last term on the r.h.s.~vanishes in the
quasi-particle limit. Up to now the role of this term in an off-shell transport theory
is not clear. However, once the equations of motion for test particles will have been 
written down the meaning of this term will become obvious. 

Let us now come back to the more general case of arbitrary width. Both equations
(\ref{eq:anasless},\ref{eq:gradsless})
mix zeroth and first order gradient terms. We first concentrate on the zeroth order
terms:
\begin{eqnarray}
  \label{eq:zerapp1}
\left(p_0 - {\vec p^2 \over 2m} - {\rm Re}\bar \Sigma^{\rm ret} \right) S^< &=&
\pm i \bar\Sigma^< \,{\rm Re}\bar D^{\rm ret} + o(\partial_X)   \,, \\
  \label{eq:zerapp2}
\Gamma \, S^< &=& \pm i \bar\Sigma^< \,{\cal A} + o(\partial_X) \,.
\end{eqnarray}
Since we already know expressions for ${\rm Re}\bar D^{\rm ret}$ and ${\cal A}$ we
insert (\ref{eq:solspec},\ref{eq:solred}) and obtain
\begin{eqnarray}
  \label{eq:nts1}
\left(p_0 - {\vec p^2 \over 2m} - {\rm Re}\bar \Sigma^{\rm ret} \right) S^< &=&
\pm i \bar\Sigma^< \,
{ p_0 - {\vec p^2 \over 2m}-{\rm Re}\bar \Sigma^{\rm ret}   \over 
\left( 
p_0 - {\vec p^2 \over 2m}-{\rm Re}\bar \Sigma^{\rm ret} 
\right)^2
+ {1 \over 4} \Gamma^2   }   + o(\partial_X)  \,, \\
  \label{eq:nts2}
\Gamma \, S^< &=& \pm i \bar\Sigma^< \,
{ \Gamma  \over 
\left( 
p_0 - {\vec p^2 \over 2m}-{\rm Re}\bar \Sigma^{\rm ret} 
\right)^2
+ {1 \over 4} \Gamma^2   }    + o(\partial_X)  \,.
\end{eqnarray}
At first glance these two equations seem to contain the same information in zeroth
order gradient expansion:
\begin{equation}
  \label{eq:slesssigless}
S^< = {\pm i \bar\Sigma^< \over 
\left( 
p_0 - {\vec p^2 \over 2m}-{\rm Re}\bar \Sigma^{\rm ret} 
\right)^2
+ {1 \over 4} \Gamma^2   }   + o(\partial_X)  
\end{equation}
obtained from (\ref{eq:nts1}) by dividing by 
$(p_0 - {\vec p^2 \over 2m}-{\rm Re}\bar \Sigma^{\rm ret})$ and from (\ref{eq:nts2})
by dividing by $\Gamma$. However, we have implicitly assumed in these steps that
both factors are not too small, i.e.
\begin{eqnarray}
  \label{eq:odx1}
{ 1 \over p_0 - {\vec p^2 \over 2m}-{\rm Re}\bar \Sigma^{\rm ret} } \,o(\partial_X) &=&
o(\partial_X)   \,, \\
  \label{eq:odx2}
{ 1 \over \Gamma} \, o(\partial_X) &=&  o(\partial_X)   \,.
\end{eqnarray}
This can only be true if we are {\it not} in the quasi-particle regime. There, the 
width is very small. Therefore, e.g.~(\ref{eq:odx2}) might be invalid. Fortunately
we are here interested in the case where the width is finite.\footnote{Nonetheless, in 
the end we want to have an equation which is also valid for small width, i.e.~shows
the correct quasi-particle limit. Indeed, as we shall see below, this requirement
is fulfilled. Since a transport equation for the quasi-particle limit is well-known
we can treat in our derivation the case where the width is large.}
Thus, (\ref{eq:slesssigless}) holds and provides information about the size of a 
specific combination of Green functions and self-energies:
\begin{equation}
  \label{eq:isc}
\pm i \bar\Sigma^< -  d\, S^<
= o(\partial_X)                 
\end{equation}
with
\begin{equation}
  \label{eq:defd}
d = \left( 
p_0 - {\vec p^2 \over 2m}-{\rm Re}\bar \Sigma^{\rm ret} 
\right)^2
+ {1 \over 4} \Gamma^2    \,.
\end{equation}
Note that $d$ is the denominator of ${\cal A}$ as well as ${\rm Re}\bar D^{\rm ret}$.
Since all our consideration are based on the assumption that space-time derivatives
are small we conclude that the l.h.s.~of (\ref{eq:isc}) also has to be small. This is
a consistency condition which might be checked in actual calculations. Since we
keep terms linear in the gradients the l.h.s.~of (\ref{eq:isc}) in general cannot be 
neglected. However, when it appears in gradients we can neglect such a combination, 
i.e.~for arbitrary $B$:
\begin{eqnarray}
  \label{eq:lastapprox}
[\pm i\bar\Sigma^<,\,B] = [d\, S^< , \, B] 
+ \underbrace{[\pm i\bar\Sigma^< - d\, S^< ,\,B] }_{o(\partial_X^2)}
= [d\, S^< , \, B]
\end{eqnarray}
where we have neglected the gradient which is effectively of second order in the gradient
expansion \cite{BotMal}. Therefore, it turns out to be consistent with the gradient
expansion up to (including) first order to replace $\pm i\bar\Sigma^<$ in the first order
gradients in (\ref{eq:anasless},\ref{eq:gradsless}) by $d\, S^< $. Then, after some 
straightforward manipulations we find that both equations reduce to the same transport
equation (cf.~\cite{IKV99})
\begin{equation}
  \label{eq:fineomsless}
{1 \over 2} \Gamma {\cal A} \,
[p_0 - {\vec p^2 \over 2m}-{\rm Re}\bar \Sigma^{\rm ret}, \, S^<]
- {1 \over 2} {\cal A} \,
[ \Gamma, \, (p_0 - {\vec p^2 \over 2m}-{\rm Re}\bar \Sigma^{\rm ret})S^<]
= \Gamma \, S^< \mp i \bar\Sigma^< {\cal A}  \,.
\end{equation}
With this equation we have reached our goal to derive in first order gradient 
expansion consistent equations for the three quantities $S^<$, ${\cal A}$, and 
${\rm Re}\bar D^{\rm ret}$. These equations are given by (\ref{eq:fineomsless}),
(\ref{eq:solspec}), and (\ref{eq:solred}), respectively. They have
also been derived in \cite{IKV99,CJ99}. It is interesting to note
that only the equation for $S^<$ is a differential equation. 

Let us recall again that
the derivation of this latter equation has been obtained in a two-step procedure.
The first step was the formal gradient expansion yielding {\it two} equations for
$S^<$, namely (\ref{eq:anasless}) and (\ref{eq:gradsless}). The second step was to 
realize that a special combination (\ref{eq:isc}) of self-energies and Green functions 
is effectively of first order in the gradient expansion. Without that second step
the two equations (\ref{eq:anasless},\ref{eq:gradsless}) would not yield identical
results and it would be unclear which equation one should use to evolve $S^<$ in time.
Of course, it is interesting to find out for which physical situations (\ref{eq:isc}) is
realized since in turn this indicates where the gradient expansion is justified. 
For that purpose we rewrite condition (\ref{eq:isc}) in the way as it appears
in the traditional transport equation (\ref{eq:gradsless}):
\begin{equation}
\label{eq:isc2}
\Gamma \, S^< \mp i \bar\Sigma^< \,{\cal A} 
= \Gamma \,{\cal A} \,
\left( {S^< \over {\cal A}} - {\pm i \bar\Sigma^< \over \Gamma} \right) 
= o(\partial_X)   \,.
\end{equation}
Obviously, there are two cases when this special combination of self-energies and 
two-point functions is small. The first case is the traditional quasi-particle regime.
There the width $\Gamma$ is small and from 
\begin{equation}
\Gamma = o(\partial_X)   
\end{equation}
one might justify the gradient expansion.\footnote{Naively one might suspect that the
spectral function becoming sharply peaked invalidates the given argument in the
region of interest, i.e.~near the on-shell point. However, in this case one should
rather compare energy integrated expressions. Then upon integration over the energy
the spectral function in front of the brackets in (\protect\ref{eq:isc2}) is basically 
replaced by one.} The second case, relevant for our purpose of describing states with
potentially large width, concerns situations where the difference of the fractions in the
brackets in (\ref{eq:isc2}) is small. Actually these two fractions define off-shell
generalizations of phase-space densities, one given by the two-point functions 
\begin{equation}
n(X,p) =  {S^<(X,p) \over {\cal A}(X,p) } 
\end{equation}
and one demanded by the self-energies
\begin{equation}
n_{\Sigma}(X,p) = {\pm i \bar\Sigma^<(X,p) \over \Gamma(X,p) }   \,.
\end{equation}
It is the difference of these two functions which in general determines the time 
evolution of the system under consideration (see also e.g.~\cite{henning,GrLe98,IKV99}).
The smallness of this difference is an alternative possible justification of the
gradient expansion if the width of the states is not small. 
Indeed, this difference is small near thermal equilibrium since there the collisional
self-energy is given by \cite{KB}
\begin{equation}
  \label{eq:setheq}
\pm i \bar\Sigma^<(p) = n_{\rm B,F}(p_0) \, \Gamma(p) 
\end{equation}
while $S^<$ is given in (\ref{eq:thermal}). This shows that both $n$ and
$n_{\Sigma}$ approach the same thermal distribution. The fact that (\ref{eq:isc2}) is 
fulfilled near thermal equilibrium does {\it not} automatically imply that it is not 
fulfilled outside of this regime, i.e.~far away from thermal
equilibrium. Whether (\ref{eq:isc2}) is realized or not for a given physical situation
has to be checked in actual calculations. If it was violated the consequence would be
that the process at hand cannot be described by first-order gradient expanded equations.

The transport equation (\ref{eq:fineomsless}) is the main result of this section.
To gain more insight into this equation we first rewrite
the r.h.s.~in a more common form. Using (\ref{eq:defspec},\ref{eq:defgam}) we get
\begin{equation}
  \label{eq:collcom}
\Gamma \, S^< \mp i \bar\Sigma^< {\cal A} = i \bar\Sigma^> S^< \mp i \bar\Sigma^< S^>
\end{equation}
which is the well-known collision term, the difference between loss and gain terms.
In the test particle description we are aiming at, the l.h.s.~of (\ref{eq:fineomsless})
describes the propagation of the test particles.
For a first interpretation of these terms on the l.h.s.~we study
the quasi-particle limit. In spite of the fact that the derivation of 
(\ref{eq:fineomsless}) is 
doubtful for infinitesimally small width (as outlined above) we will find that 
(\ref{eq:fineomsless}) nonetheless reproduces the correct transport equation for the
quasi-particle regime. In this regime one may define an on-shell phase-space density
$n$ via
\begin{equation}
  \label{eq:ospsd}
S^<(X,p) = n(X;p_0=\epsilon(X,\vec p),\vec p) \, {\cal A}(X,p)   
\end{equation}
where $\epsilon$ is the on-shell energy, i.e.~the solution $p_0 = \epsilon$ of the 
equation
\begin{equation}
  \label{eq:onshen}
p_0 - {\vec p^2 \over 2m}-{\rm Re}\bar \Sigma^{\rm ret}(X,p) = 0  \,.
\end{equation}
In this way (\ref{eq:fineomsless}) can be transformed into a transport equation for
$n$. We integrate the transport equation over $p_0$ and note that for vanishing width
both ${\cal A}$ and ${1 \over 2} \Gamma {\cal A}^2$ approach the same $\delta$-function:
\begin{equation}
  \label{eq:quasipspec}
\left.
{ {\cal A} \atop  {1 \over 2} \Gamma {\cal A}^2} 
\right\}
\to  
2 \pi \,\delta(p_0 - {\vec p^2 \over 2m}-{\rm Re}\bar \Sigma^{\rm ret}) 
= 2 \pi \, z \, \delta(p_0 -\epsilon) 
\quad \mbox{for} \quad \Gamma \to 0
\end{equation}
with the renormalization factor \cite{BotMal}
\begin{equation}
  \label{eq:defrenormquasi}
z(X,\vec p) = \left. \left( 
1 - {\partial {\rm Re}\bar \Sigma^{\rm ret}(X,p) \over \partial p_0}
\right)^{-1} \,\right\vert_{p_0 = \epsilon}      \,.
\end{equation}
The collision term, i.e.~the r.h.s.~of (\ref{eq:fineomsless}), becomes after integrating
over $p_0$:
\begin{equation}
  \label{eq:collonsh}
\left. \vphantom{\int} 
-z \left(\pm i \bar\Sigma^< (1 \pm n) -i\bar\Sigma^> n \right)
\,\right\vert_{p_0 = \epsilon}     \,.
\end{equation}
The first term on the l.h.s.~of (\ref{eq:fineomsless}) yields the Vlasov part (see also
(\ref{eq:driftstand})):
\begin{equation}
  \label{eq:vlasonsh}
\left. \vphantom{\int} [p_0 -\epsilon , \, n] \,\right\vert_{p_0 = \epsilon}     \,.
\end{equation}
Note that the $z$-factor is implicitly incorporated here on account of
\begin{equation}
  \label{eq:zimpl}
\vec \nabla_X \epsilon = \left. \vphantom{\int} 
z \vec \nabla_X {\rm Re}\bar \Sigma^{\rm ret} \,\right\vert_{p_0 = \epsilon}  
\end{equation}
and corresponding equations for derivatives with respect to momenta. The second term on 
the l.h.s.~of (\ref{eq:fineomsless}) is a genuine off-shell 
contribution. After performing the integration over $p_0$ this term vanishes for 
$\Gamma \to 0$. The origin of this term can be traced back to the last term on the 
r.h.s.~of (\ref{eq:gradsless}). As already emphasized the role of this term in an 
off-shell transport theory still requires a proper interpretation which we will give 
below in the test particle picture. To summarize, we have recovered from 
(\ref{eq:fineomsless}) the traditional Boltzmann equation
in the quasi-particle limit. 

In the next sections we will elaborate on the more general
case of non-vanishing width. Among other things we will address two questions: 
1.~Is the particle number still a conserved quantity when the transport equation 
(\ref{eq:fineomsless}) is used to describe the time evolution of $S^<$? 
2.~What is the role of the off-shell contribution in (\ref{eq:fineomsless})?


\section{Effective particle number}
\label{sec:ppn}

We are aiming at a test particle representation for the transport process described
by (\ref{eq:fineomsless}). For a test particle representation 
to make sense it is mandatory that the represented quantity is conserved as outlined
in Sec.~\ref{sec:simple}. 
Suppose that we have chosen the self-energies such (conserving approximation
\cite{KB,BotMal,IKV99}) that the particle number
(\ref{eq:partnumbexact}) is conserved for the full quantum theory defined by
the exact equations of motion (\ref{eq:exeom1},\ref{eq:exeom2}). This does
not necessarily mean that also the transport equation (\ref{eq:fineomsless}) conserves 
this quantity. Indeed, as we will argue below this is not the case. 
The reason is simply that by the coarse graining process carried out by 
the gradient expansion the part of $S^<(X,p)$ which is highly oscillating in $X$ has 
been neglected. Of course, the temporal deviation in the particle number is effectively
of higher order in the gradient expansion since up to first order 
the exact and the approximate equations of motion carry the same information.
However, it would be very unpleasant if one had to represent a quantity by test 
particles which is not fully conserved on the level of approximations. Thus, after
showing that the particle number is not fully conserved we shall search for another
quantity which is exactly conserved by the transport equation. 

To study the time evolution of the particle number (\ref{eq:partnumbexact}) we divide
(\ref{eq:fineomsless}) by ${1 \over 2}\Gamma{\cal A}$ and integrate over the space
coordinates and the four-momentum. All gradients with respect to $\vec x$ and $\vec p$
vanish by partial integration (and neglecting surface terms). Finally we get
\begin{eqnarray}
{d \over dt} N(t) 
&=& {d \over dt} \int\!\! d^3\!x \int\!\!{d^4\!p \over (2\pi)^4} S^<(t,\vec x;p) 
\nonumber \\
  \label{eq:notcons}
&=& {d \over dt} 
\int\!\! d^3\!x \int\!\!{d^4\!p \over (2\pi)^4} S^< K
+ \int\!\! d^3\!x \int\!\!{d^4\!p \over (2\pi)^4} {2 \over \Gamma {\cal A}}
\left( \pm i \bar\Sigma^< S^> - i \bar\Sigma^> S^< \right)
\end{eqnarray}
with
\begin{equation}
  \label{eq:defK}
K(X,p) = {\partial {\rm Re}\bar \Sigma^{\rm ret}(X,p) \over \partial p_0} +
{p_0 - {\vec p^2 \over 2m}-{\rm Re}\bar \Sigma^{\rm ret}(X,p)  \over \Gamma(X,p) }
{\partial \Gamma(X,p) \over \partial p_0}   \,.
\end{equation}
Note that the quantity $1-K$ is the off-shell generalization of the inverse 
quasi-particle renormalization factor $1/z$ as defined in (\ref{eq:defrenormquasi}).

Now we have to specify the self-energies. To keep things simple we choose a two-body
interaction 
\begin{equation}
  \label{eq:twobodyham}
{1 \over 2} \int \!\! d^3\!x \, d^3\!x' \, 
\psi^\dagger(t,\vec x) \, \psi^\dagger(t,\vec x') \, v(\vert \vec x -\vec x' \vert) \,
\psi(t,\vec x') \, \psi(t,\vec x)
\end{equation}
and evaluate the collisional self-energies in the Born approximation \cite{KB,Da84a}
\begin{eqnarray}
i \bar \Sigma^>(X,p) &=& 
\int\!\! {d^4\!p_1 \over (2\pi)^4}{d^4\!p_2 \over (2\pi)^4}{d^4\!p_3 \over (2\pi)^4} \,
(2\pi)^4 \delta^{(4)}(p+p_1-p_2-p_3) 
\nonumber \\
  \label{eq:born1}
&& \times {1 \over 2} 
\left(\bar v(\vec p -\vec p_2) \pm \bar v(\vec p -\vec p_3) \right)^2
S^<(X,p_1) \, S^>(X,p_2) \, S^>(X,p_3)  \,,   \\
\pm i \bar \Sigma^<(X,p) &=& 
\int\!\! {d^4\!p_1 \over (2\pi)^4}{d^4\!p_2 \over (2\pi)^4}{d^4\!p_3 \over (2\pi)^4} \,
(2\pi)^4 \delta^{(4)}(p+p_1-p_2-p_3) 
\nonumber \\
  \label{eq:born2}
&& \times {1 \over 2} 
\left(\bar v(\vec p -\vec p_2) \pm \bar v(\vec p -\vec p_3) \right)^2
S^>(X,p_1) \, S^<(X,p_2) \, S^<(X,p_3)  \,.
\end{eqnarray}
For a more elaborate treatment of self-energies we refer to \cite{IKV99}.

The two contributions on the r.h.s.~of (\ref{eq:notcons}) show completely different
structures. Especially we note that the first term belongs to the 
part of the transport equation which describes the propagation of modes while the
second term comes from the collisional part.
It is hard to conceive that these two terms can cancel each other. Indeed, we have not
managed to rearrange these terms in a way that one can get an idea why the r.h.s.~of
(\ref{eq:notcons}) should vanish. We therefore strongly conjecture that $N(t)$ is 
{\it not} conserved by the transport equation (\ref{eq:fineomsless}). At this stage a 
comparison to the approach of \cite{BotMal} is in order. There terms like the first one
on the r.h.s.~of (\ref{eq:notcons}) are neglected since it contains a double derivative
--- one with respect to $t$ and one with respect to $p_0$. We cannot neglect
such contributions since we have only assumed that space-time derivatives are small.
We have not made any assumptions about the smallness of derivatives with respect to 
momenta. Even more important, we are not aiming at a quantity which is nearly conserved 
by the transport equation up to terms which are of higher order in the gradients. To
represent the transport equation by test particles we need a quantity which is
exactly conserved. Of course, this difference is somewhat subtle since e.g.~the 
particle number is exactly conserved by the full quantum mechanical equation of motion.
However, it is not the latter equation we want to solve by a test particle ansatz but 
the transport equation obtained by gradient expansion.

We will now construct a quantity which is conserved by (\ref{eq:fineomsless}). 
The first step is to get rid of the collisional part. 
By inspecting (\ref{eq:born1},\ref{eq:born2}) we observe that the following integral 
vanishes \cite{KB,BotMal}:
\begin{equation}
  \label{eq:collvan}
\int\!\!{d^4\!p \over (2\pi)^4} 
\left( \pm i \bar\Sigma^< S^> - i \bar\Sigma^> S^< \right)   = 0  \,.
\end{equation}
This suggests to directly integrate (\ref{eq:fineomsless}) over momenta and space
coordinates instead of dividing it by ${1 \over 2}\Gamma{\cal A}$ first. Indeed, for
the quantity 
\begin{equation}
  \label{eq:defstilde}
{\cal S} = {1 \over 2}\Gamma{\cal A} S^<
\end{equation}
we derive from (\ref{eq:fineomsless}) the following equation of motion 
(see Appendix \ref{app:eomstilde}):
\begin{equation}
  \label{eq:eomstilde}
[p_0 - {\vec p^2 \over 2m}-{\rm Re}\bar \Sigma^{\rm ret}, \, {\cal S}]
- {1 \over \Gamma}  \,
[ \Gamma, \, (p_0 - {\vec p^2 \over 2m}-{\rm Re}\bar \Sigma^{\rm ret}){\cal S}]
= i \bar\Sigma^> S^< \mp i \bar\Sigma^< S^>  \,.
\end{equation}
Thus we find
\begin{equation}
  \label{eq:conslaw}
{d \over dt} \int\!\! d^3\!x \int\!\!{d^4\!p \over (2\pi)^4} {\cal S}
= {d \over dt} \int\!\! d^3\!x \int\!\!{d^4\!p \over (2\pi)^4} {\cal S} K  \,.
\end{equation}
We conclude that the space and momentum integral of the quantity 
\begin{equation}
  \label{eq:defpseudon}
\tilde S^< = {\cal S} \, (1-K) = {1 \over 2} \Gamma {\cal A} S^< \, (1-K)
\end{equation}
is conserved by the transport equation (\ref{eq:fineomsless}): 
\begin{equation}
  \label{eq:defntilde}
{d \over dt} \tilde N(t) = 
{d \over dt} \int\!\! d^3\!x \int\!\!{d^4\!p \over (2\pi)^4} \tilde S^<(t,\vec x;p) 
= \int\!\! d^3\!x \int\!\!{d^4\!p \over (2\pi)^4} 
\left( \pm i \bar\Sigma^< S^> - i \bar\Sigma^> S^< \right)= 0  \,.
\end{equation}
In the following we will refer to $\tilde N$ as the effective particle number. 
In the quasi-particle
limit, $\Gamma \to 0$, we find
\begin{equation}
  \label{eq:pseudoquasi}
\tilde S^<(X,p) \to n(X;\epsilon,\vec p) \, 2\pi \, \delta(p_0-\epsilon)
\end{equation}
where we have used (\ref{eq:ospsd},\ref{eq:quasipspec}) and
\begin{equation}
  \label{eq:relzk}
1-K  \to z^{-1}   \,.
\end{equation}
Thus, in the quasi-particle limit there remains an interesting difference between
\begin{equation}
  \label{eq:slquasi}
S^<(X,p) \to n(X;\epsilon,\vec p) \, 2\pi \, z \, \delta(p_0-\epsilon)
\end{equation}
and $\tilde S^<$. The quantity which "counts" the quasi-particles is indeed $\tilde S^<$
while the renormalization factor $z$ appears in $S^<$ which takes into account that
there is some strength in the spectral function (far) away from the quasi-particle pole.

\begin{figure}
\centerline{\psfig{figure=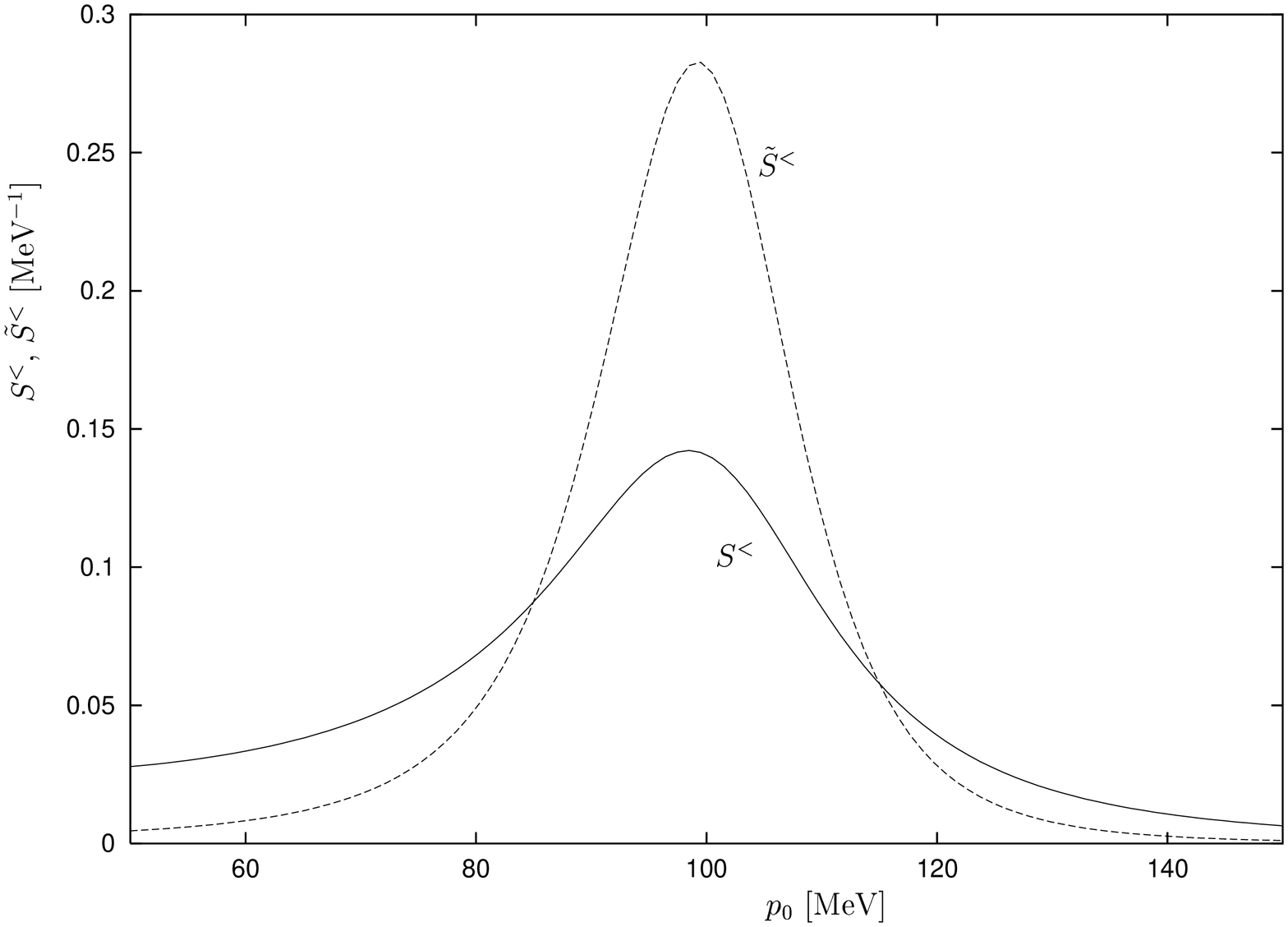,height=12cm,angle=-90}}
\caption{$S^<_{\rm th}$ (full) and $\tilde S^<_{\rm th}$ (dashed) as functions of the 
energy $p_0$ for
the case of bosons. The width is chosen to be constant, $\Gamma = 30\,$MeV. The
temperature is $T = 150\,$MeV and the kinetic energy ${\vec p^2 \over 2m} = 100\,$MeV. 
See main text for more details.}
\label{fig:stherm}
\end{figure}

As a second limiting case we evaluate $\tilde S^<$ for thermal equilibrium. 
Using (\ref{eq:thermal}) we get
\begin{equation}
  \label{eq:Stherm}
\tilde S^<_{\rm th}(p) 
= n_{\rm B,F}(p_0) \, {1 \over 2} \Gamma(p) {\cal A}^2(p)\,  (1-K(p))  \,.
\end{equation}
For constant width and vanishing ${\rm Re}\bar \Sigma^{\rm ret}$ the quantities
$S^<_{\rm th}$ and $\tilde S^<_{\rm th}$ are depicted in Fig.~\ref{fig:stherm}.
Obviously the function ${1 \over 2} \Gamma {\cal A}^2$ is more strongly 
peaked than ${\cal A}$ (also cf.~\cite{SL95}). The asymmetric forms are caused
by the Bose enhancement. 

Comparison of (\ref{eq:Stherm}) with (\ref{eq:thermal}) suggests to introduce 
--- also for non-equilibrium situations --- an effective spectral function
\begin{equation}
  \label{eq:defpsspec}
\tilde {\cal A}(X,p) = {1 \over 2} \Gamma(X,p) \, {\cal A}^2(X,p)\,  (1-K(X,p))   \,.
\end{equation}
Indeed, as will be shown in Appendix \ref{app:psspec} the effective spectral function
is also normalized (cf.~(\ref{eq:normspec})). 
Thus we have a correspondence of the quantities $S^<$, $N$, and ${\cal A}$ of the full
theory with the quantities $\tilde S^<$, $\tilde N$, and $\tilde {\cal A}$ of the 
coarse grained theory obtained by gradient expansion. However, a word of caution is
in order here. For the interpretation of $\tilde {\cal A}$ as an effective spectral 
function
(and also of $\tilde N$ as an effective particle number!) it is mandatory that 
$\tilde {\cal A}$ is always non-negative. This may not be the case
for arbitrary self-energies. Since $\Gamma$ is always positive one has to make sure
in actual calculations that 
$1-K$ is non-negative. Otherwise a test particle ansatz would not make any sense.
Finally we note that $\tilde {\cal A}$ was also introduced in \cite{IKV99} as the
spectral information which enters the entropy density.


\section{Equations of motion for test particles} \label{sec:tpeom}

The next step is the test particle ansatz for $\tilde S^<$ and the derivation of the
equations of motion for the test particles from the transport equation 
(\ref{eq:fineomsless}):
\begin{equation}
  \label{eq:testpa}
\tilde S^<(t,\vec x;p) \sim 
\sum\limits_i \delta^{(3)}(\vec x - \vec x_i(t)) \, \delta(p_0 - E_i(t)) \,
\delta^{(3)}(\vec p - \vec p_i(t))   \,.
\end{equation}
Note that the energy $E_i$ of the test particle $i$ is a free coordinate, not restricted
by a mass-shell condition. 
Instead of (\ref{eq:fineomsless}) we use the equivalent equation (\ref{eq:eomstilde}).
While the r.h.s.~describes the collisions the l.h.s.~will give the propagation
of the test particles between collisions. To get the equation of motion for a single 
test particle {\it between two collisions} we have to insert
\begin{equation}
  \label{eq:testtild}
{\cal S}(t,\vec x;p) = { \tilde S^<(t,\vec x;p) \over 1-K(t,\vec x;p) } 
\sim \sum\limits_i {1 \over 1-K(t,\vec x_i;E_i,\vec p_i) } 
\delta^{(3)}(\vec x - \vec x_i(t)) \, \delta(p_0 - E_i(t)) 
\, \delta^{(3)}(\vec p - \vec p_i(t)) 
\end{equation}
into
\begin{equation}
  \label{eq:eomstildenocoll}
[p_0 - {\vec p^2 \over 2m}-{\rm Re}\bar \Sigma^{\rm ret}, \, {\cal S}]
- {1 \over \Gamma}  \,
[ \Gamma, \, (p_0 - {\vec p^2 \over 2m}-{\rm Re}\bar \Sigma^{\rm ret}){\cal S}]
= 0 \,.
\end{equation}
In general this yields an equation of the following type
\begin{eqnarray}
\sum\limits_i \left( 
a(t,\vec x_i;E_i,\vec p_i) + b(t,\vec x_i;E_i,\vec p_i) \partial_{p_0} + 
\vec c(t,\vec x_i;E_i,\vec p_i) \vec\nabla_x 
+ \vec d(t,\vec x_i;E_i,\vec p_i) \vec\nabla_p 
\right) &&   \nonumber \\
  \label{eq:gentp}
\otimes \delta^{(3)}(\vec x - \vec x_i) \, \delta(p_0 - E_i) \,
\delta^{(3)}(\vec p - \vec p_i)   &=& 0   \,.
\end{eqnarray}
A solution for this equation is obtained by demanding that all coefficient functions
$a$, $b$, $\vec c$, and $\vec d$ have to vanish. In general this gives eight equations
for the seven test particle coordinates $E_i$, $\vec x_i$, and $\vec p_i$. 
If and only if the quantity (in our case $\tilde S^<$) which is represented by test 
particles corresponds to a conserved quantity (here $\tilde N$) then the coefficient
$a$ in (\ref{eq:gentp}) vanishes and one only has to fulfill seven equations instead of
eight. This is the reason why we have insisted to find an
exactly conserved quantity. Otherwise one would have to deal with an overdetermined
system of equations. Neglecting one of the obtained equations would not yield a solution
for the transport equation one actually wants to solve (cf.~the corresponding 
discussion in Sec.~\ref{sec:simple}). For the 
case at hand we get the following equations of motion for the test particles
\begin{eqnarray}
  \label{eq:tpeom1}
0 &\stackrel{!}{=}& 
- \dot E_i + 
{1 \over 1-K} \left(
\partial_t {\rm Re}\bar \Sigma^{\rm ret}
+ {E_i - { \vec p_i^2 \over 2m} - {\rm Re}\bar \Sigma^{\rm ret}  \over \Gamma } 
\partial_t \Gamma 
\right)   
= b(t,\vec x_i;E_i,\vec p_i)   \,,  \\
  \label{eq:tpeom2}
0 &\stackrel{!}{=}&  - \dot{\vec x_i} 
+ {1 \over 1-K} \left(
{\vec p_i \over m} + \vec\nabla_{p_i} {\rm Re}\bar \Sigma^{\rm ret} +
{E_i - { \vec p_i^2 \over 2m} - {\rm Re}\bar \Sigma^{\rm ret}  \over \Gamma } 
\vec\nabla_{p_i} \Gamma
\right)
= \vec c(t,\vec x_i;E_i,\vec p_i)   \,,  \\
  \label{eq:tpeom3}
0 &\stackrel{!}{=}&   - \dot{\vec p_i} 
- {1 \over 1-K} \left(
\vec\nabla_{x_i} {\rm Re}\bar \Sigma^{\rm ret} +
{E_i - { \vec p_i^2 \over 2m} - {\rm Re}\bar \Sigma^{\rm ret}  \over \Gamma } 
\vec\nabla_{x_i} \Gamma
\right)
= \vec d(t,\vec x_i;E_i,\vec p_i)   \,.
\end{eqnarray}
Newton's equations of motion are obtained by neglecting $K$, i.e.~putting
the renormalization factor $1/(1-K)$ to 1, and disregarding all terms where the width
$\Gamma$ enters. In this limiting case one observes that the energy of a test particle 
is only 
changed if the real part of the self-energy, i.e.~the classical potential, depends
explicitly on time. If the potential shows an explicit momentum dependence the
classical expression for the velocity is modified accordingly. For finite width but
neglecting the energy and momentum dependence of the self-energies 
(\ref{eq:tpeom1}-\ref{eq:tpeom3}) basically reduce to the test particle equations 
presented in \cite{CJ99}. The only remaining difference is that there a relativistic 
system is studied.

Let us now come back to the full equations of motion (\ref{eq:tpeom1}-\ref{eq:tpeom3}).
Obviously the terms which involve the width yield genuine off-shell contributions to all
equations of motion since on the mass-shell the combination 
$E_i - { \vec p_i^2 \over 2m} - {\rm Re}\bar \Sigma^{\rm ret}$ vanishes by definition
(cf.~(\ref{eq:onshen})). 
It is interesting to calculate the time evolution of the off-shellness of a test
particle defined by
\begin{equation}
  \label{eq:defoffsn}
\Delta E_i(t,\vec x_i;E_i,\vec p_i) = 
E_i - { \vec p_i^2 \over 2m} - 
{\rm Re}\bar \Sigma^{\rm ret} (t,\vec x_i;E_i,\vec p_i)    \,.
\end{equation}
From the equations of motion (\ref{eq:tpeom1}-\ref{eq:tpeom3}) we find 
\begin{equation}
  \label{eq:eomoffsn}
{d \over dt} \Delta E_i = {\Delta E_i \over \Gamma} {d \over dt}\Gamma   \,.
\end{equation}
This equation has also been presented in \cite{CJ99} and in \cite{EM99}. To the
best of our knowledge the full set of equations of motion 
(\ref{eq:tpeom1}-\ref{eq:tpeom3}) has never been derived before. Obviously the time
evolution of the off-shellness is caused by the respective last term in the brackets 
on the l.h.s.~of (\ref{eq:tpeom1}-\ref{eq:tpeom3}). These contributions can be traced 
back to the second term on the l.h.s.~of (\ref{eq:fineomsless}). This clarifies the 
meaning of the latter term for an off-shell transport theory: It provides the time 
evolution of the off-shellness. Actually this off-shell contribution in the transport
equation (\ref{eq:fineomsless}) is caused by the respective last term on the r.h.s.~of
(\ref{eq:anasless}) and (\ref{eq:gradsless}) via the replacement (\ref{eq:lastapprox}). 
Without that latter replacement the meaning of these terms in 
(\ref{eq:anasless},\ref{eq:gradsless}) would have been completely unclear. 

The equations of motion (\ref{eq:tpeom1}-\ref{eq:tpeom3}) obeyed by the test particles 
between collisions are 
the main result of this section. It is useful to show that they reproduce the correct 
quasi-particle limit. The quasi-particle energy of a test particle is defined as the
solution $\epsilon_i = \epsilon_i(t,\vec x_i;\vec p_i)$ of the equation
\begin{equation}
  \label{eq:onshtp}
\epsilon_i - { \vec p_i^2 \over 2m} - 
{\rm Re}\bar \Sigma^{\rm ret} (t,\vec x_i;\epsilon_i,\vec p_i)  = 0   \,.
\end{equation}
Differentiating (\ref{eq:onshtp}) e.g.~with respect to $\vec p_i$ yields 
(also cf.~(\ref{eq:zimpl}))
\begin{equation}
  \label{eq:onedp}
\vec\nabla_{p_i} \epsilon_i = {1 \over 1 - K(t,\vec x_i;\epsilon_i,\vec p_i)}
\left(
{\vec p_i \over m} + 
\vec\nabla_{p_i} {\rm Re}\bar \Sigma^{\rm ret}(t,\vec x_i;\epsilon_i,\vec p_i)
\right)  \,.
\end{equation}
Thus we find the expected relations
\begin{eqnarray}
  \label{eq:eomsimple1}
\dot{\vec x_i} &=& \vec \nabla_{p_i} \epsilon_i    \,,  \\
  \label{eq:eomsimple2}
\dot{\vec p_i} &=& -\vec \nabla_{x_i} \epsilon_i    \,.
\end{eqnarray}
The time evolution of the on-shell energy is governed by
\begin{equation}
  \label{eq:onssimple}
\dot \epsilon_i = {1 \over 1-K(t,\vec x_i;\epsilon_i,\vec p_i)} 
\partial_t {\rm Re}\bar \Sigma^{\rm ret}(t,\vec x_i;\epsilon_i,\vec p_i) 
= z(t,\vec x_i;\epsilon_i,\vec p_i) \,
\partial_t {\rm Re}\bar \Sigma^{\rm ret}(t,\vec x_i;\epsilon_i,\vec p_i)   \,.
\end{equation}

Finally we comment on the evaluation of the self-energies which enter the equations
of motion (\ref{eq:tpeom1}-\ref{eq:tpeom3}) and of course also the collisional part
on the r.h.s.~of the transport equation (\ref{eq:fineomsless}). Except for the 
Hartree-Fock self-energy which is comparatively easy to incorporate in transport 
calculations all self-energy contributions can be traced back to the determination of 
$\pm i\bar\Sigma^<$ and $i\bar\Sigma^>$ via the relations 
(\ref{eq:defgam},\ref{eq:dispsigret},\ref{eq:defK}). In the Born approximation to
two-body collisions these self-energies are given in (\ref{eq:born1},\ref{eq:born2}). 
These quantities should be expressed in terms of $\tilde S^<$ since the latter quantity
is represented by test particles according to (\ref{eq:testpa}). We find for the
Born rates
\begin{eqnarray}
i \bar \Sigma^>(X,p) &=& 
\int\!\! {d^4\!p_1 \over (2\pi)^4}{d^4\!p_2 \over (2\pi)^4}{d^4\!p_3 \over (2\pi)^4} \,
(2\pi)^4 \delta^{(4)}(p+p_1-p_2-p_3) \, {1 \over 2} 
\left(\bar v(\vec p -\vec p_2) \pm \bar v(\vec p -\vec p_3) \right)^2
\nonumber \\
  \label{eq:born1test}
&& \times 
{\tilde S^<_1 \over {1\over 2}\Gamma_1\,{\cal A}_1\, (1-K_1)}
\, 
{\tilde {\cal A}_2 \pm \tilde S^<_2 \over {1\over 2}\Gamma_2\,{\cal A}_2 \, (1-K_2)} 
\, 
{\tilde {\cal A}_3 \pm \tilde S^<_3 \over {1\over 2}\Gamma_3\,{\cal A}_3 \, (1-K_3)} 
\,,   \\
\pm i \bar \Sigma^<(X,p) &=& 
\int\!\! {d^4\!p_1 \over (2\pi)^4}{d^4\!p_2 \over (2\pi)^4}{d^4\!p_3 \over (2\pi)^4} \,
(2\pi)^4 \delta^{(4)}(p+p_1-p_2-p_3) \, {1 \over 2} 
\left(\bar v(\vec p -\vec p_2) \pm \bar v(\vec p -\vec p_3) \right)^2
\nonumber \\
  \label{eq:born2test}
&& \times 
{\tilde {\cal A}_1 \pm \tilde S^<_1 \over {1\over 2}\Gamma_1\,{\cal A}_1 \, (1-K_1)} 
\,
{\tilde S^<_2 \over {1\over 2}\Gamma_2\,{\cal A}_2\, (1-K_2)}
\, 
{\tilde S^<_3 \over {1\over 2}\Gamma_3\,{\cal A}_3\, (1-K_3)}
\end{eqnarray}
with $\tilde S^<_j = \tilde S^<(X,p_j)$ etc. Obviously the fact that $\tilde S^<$ instead
of $S^<$ is represented by test particles causes rather non-trivial modifications
for the collision integrals. In turn, disregarding these modifications amounts to 
solving transport equations which might be rather different from the ones one actually
wants to solve.


\section{Summary and outlook}   \label{sec:sum}

In this work we have presented a derivation of a transport equation making use of the
gradient expansion but without utilizing the commonly used quasi-particle
approximation. Therefore, the derived equations can be used for the description
of processes where the width of the involved modes might be arbitrarily 
large.\footnote{Of course, for a non-relativistic formalism to make sense one has to 
assume that kinetic and potential energy and also the width is small as compared to the
rest mass of the particle.}
In principle, the gradient expansion yields six equations for the three quantities of 
interest. We have successively shown which of these equations are redundant by various
rearrangements. In the course of this procedure it has become apparent that a specific
combination of self-energies and Green functions --- which is seemingly of zeroth order 
in the gradient expansion --- must be treated as being of first order. This 
identification (\ref{eq:isc}) is mandatory to derive a contribution to the transport
equation which finally leads to the description of the time evolution of the 
off-shellness of the test particles. As a next step a quantity --- the effective particle
number --- has been isolated which is exactly conserved by the transport equation. 
This has opened the way to a test particle representation of the
density $\tilde S^<$ which corresponds to the effective particle number. 
The equations of motion for
the test particles have been presented. Finally the Born collision rates are
evaluated within the test particle representation. 

Remarkably the effective particle number is {\it not} identical to the
particle number which is exactly conserved by the fully quantum field theoretical
equations of motion. Of course, the particle number is approximately conserved by
the transport equation since up to first order in the gradient expansion the 
approximate (transport) and the exact (quantum field theoretical) equation carry the
same information. To apply a test particle ansatz to the transport equation one needs,
however, a quantity which is exactly conserved. Nonetheless, in actual simulations
the particle number $N$ of the full theory might serve as a test for the accuracy of the
gradient expansion. If during the time evolution $N$ deviates drastically from its 
initial value the gradient expansion has to be regarded as inappropriate for the system
one wants to simulate. 

Throughout this work we have strictly distinguished between the (possibly off-shell)
propagation of the test particles between collisions on the one hand side and the binary
collisions on the other. This distinction might seem to be artificial since the width 
which allows for an off-shell propagation is caused by the collisions. Since our
equations are derived from the underlying quantum field theory in a well-defined and
controlled way we do not have to be afraid of double counting. The physical picture we 
have in mind which corresponds to our approach are collisions which are so frequent
that particles do not come back to their mass-shell before the next collision happens.
Therefore part of the collision process has to be incorporated in the propagation
by allowing the propagation of test particles with arbitrary energy. An additional
aspect which we have not touched here but is in principle straightforward to include
is the treatment of resonances with large decay width. Here it is very natural that
a propagating resonance is simulated by a bunch of test particles with variable energies.
We note that our formalism treats collisional and decay width on equal footing. 

We would like to comment briefly on the possible numerical realization of the presented 
approach. To calculate for a given space-time point the spectral function and the width 
which enter the collision integrals (\ref{eq:born1test},\ref{eq:born2test}) one has to
know the self-energies for arbitrary four-momenta. This requires the use of much more
test particles as compared to simulations which are restricted to the quasi-particle
regime. In present simulations of off-shell effects in transport theories rather ad-hoc 
recipes are in use to overcome this problem. In \cite{CJ99} the self-energies are
not determined independently for every point in space but rather from an average over
a larger volume. 
Whether this is a valid approximation to the full treatment of the self-energies
remains to be seen.
In \cite{EM99} the self-energies for a given space-time point are
parametrized by an expression adopted from the case of thermal equilibrium.
The thermal parameters are locally chosen such that a reasonable fit on the shape 
suggested by the test particles is obtained. It has been checked there that time as well
as momentum averages of the parametrized self-energies agree with the corresponding
results from the test particle representation.

Next we compare our formalism to recent other approaches. In \cite{IKV99} it has been
worked out in great detail how general self-energies have to be chosen such that 
particle number and energy are exactly conserved by the full quantum field theoretical 
Kadanoff-Baym equations. It has not been shown there that these quantities are
{\it exactly} conserved by the approximate gradient expanded equations. As we have 
outlined above we conjecture that this is indeed not the case. To some extent it is
trivial that quantities which are exactly conserved by the exact equations are 
(at least) approximately conserved by the approximate equations. Nonetheless it is
interesting to see how this approximate conservation comes about in the approximate
transport formalism \cite{BotMal}. 
In our approach, however, we had to answer a somewhat
different question. For the test particle realization we had to find a quantity
which is exactly conserved by the approximate transport equation. 

The purpose of
\cite{CJ99} and \cite{EM99} was also to present a test particle description of
off-shell transport processes. In spite of the fact that in \cite{CJ99} the transport
equation (\ref{eq:fineomsless}) was derived in basically the same way as presented here,
the approach of \cite{CJ99} is very different in spirit. There the gradient expanded 
Kadanoff-Baym equations were only used to determine the evolution of test particles
between collisions. In contrast, the information about how collisions between these
test particles have to be treated was taken from a completely different source.
Since in \cite{CJ99} it was anyway not the purpose to solve equation 
(\ref{eq:fineomsless}) it did not matter whether the test particle equations of motion
were derived from a test particle ansatz for $S^<$ or for $\tilde S^<$. 
Obviously, in the approach presented here we take the transport equation 
(\ref{eq:fineomsless}) serious as being an equation which can be obtained from the
underlying quantum field theoretical equations in a straightforward and well-controlled
manner by gradient expansion. A second difference between our approach and the one
presented in \cite{CJ99} is the fact that the energy-momentum
dependence of the self-energies was neglected for the test particle equations of motion.
Therefore, the test particle equations presented here are more general than the ones
given in \cite{CJ99}.\footnote{After finishing this work a preprint \cite{CJ992}
appeared where the authors of \cite{CJ99} generalized their relativistic test particle 
equations by also including energy-momentum dependences of the self-energies.}
Finally we studied here a non-relativistic system while in 
\cite{CJ99} the case of relativistic bosons was treated. 

In \cite{EM99} the starting point was the transport equation (\ref{eq:gradsless})
but neglecting the last term on the r.h.s.
The equation (\ref{eq:eomoffsn}) which describes the time 
evolution of the off-shellness was introduced by hand based on physical plausibility
arguments. The collision terms were treated 
in the way adopted from on-shell transport theories by simulating $S^<$ instead of
$\tilde S^<$ by test particles. 

While discussing at length the (non-)conservation of the particle number we have not
touched the issue of energy conservation. In view of the fact that it was non-trivial
to find a conserved effective particle number it might be no surprise to realize that
also the conservation of energy as defined for the full quantum field theoretical
equations becomes a problem for the gradient expanded equations. To clarify the
important question of energy conservation is beyond the scope of the present paper.
We only want to note here that the energy as defined for the full theory 
(see e.g.~\cite{KB,BotMal,IKV99}) has to be at least approximately conserved by the
transport equation. In principle this provides a check on the validity of the 
gradient expansion.

Finally we give a brief outlook on the generalization of our approach
to relativistic systems. Most of the presented formalism can be immediately generalized,
especially for scalar particles. As already mentioned above 
(after Eq.~(\ref{eq:trivial})) it might appear for fermions that the set of four 
equations for spectral function and real part of the retarded propagator derived in 
first order gradient expansion cannot be trivially reduced to two independent equations.
This might lead to new constraints for self-energies and Green functions. 
A second potential problem of a relativistic off-shell transport theory
concerns the possibility of space-like modes. In principle, the spectral function
might contain strength in the sector where the momentum is larger than the energy. 
Thus it can happen that in collisions space-like test particles are created which
travel faster than the speed of light. This is clearly an unpleasant feature of a
theory which should respect causality. For relativistic systems one therefore has to 
think about the proper treatment of the space-like part of the spectral function.


\acknowledgements I acknowledge stimulating discussions with Sascha Juchem, 
Martin Effenberger, Carsten Greiner, J\"orn Knoll, Wolfgang Cassing, and Ulrich Mosel.
I also thank Kimmo Kainulainen for pointing out to me the way to obtain the elegant 
equation (\ref{eq:trivial}).


\appendix

\section{Normalization of spectral function and effective spectral function}   
\label{app:psspec}

In this appendix we will 
prove the normalization 
of the spectral function ${\cal A}$
as given in (\ref{eq:solspec}) and the effective spectral function $\tilde {\cal A}$
as defined in (\ref{eq:defpsspec}). We implicitly assume in the following that
$\tilde {\cal A}$ is always non-negative (cf.~the remarks after (\ref{eq:defpsspec})). 
For the spectral function we find
\begin{equation}
  \label{eq:normspeccont}
\int\limits_{-\infty}^\infty \!\! {dp_0 \over 2\pi} {\cal A}(X,p) =
-2 {\rm Im} \int\limits_{-\infty}^\infty \!\! {dp_0 \over 2\pi} \bar D^{\rm ret}(X,p) \,.
\end{equation}
Since $\bar D^{\rm ret}(X,p)$ has only poles in the lower complex half-plane one can
use contour integration to evaluate the last integral. We assume that the self-energies
can be neglected for large $\vert p_0 \vert$, i.e.
\begin{equation}
  \label{eq:largep0}
\bar D^{\rm ret}(X,p) \to {1 \over p_0} \qquad \mbox{for} \quad \vert p_0 \vert \to 
\infty  \,.
\end{equation}
This allows us to write
\begin{eqnarray}
  \label{eq:normspeccont2}
\int\limits_{-\infty}^\infty \!\! {dp_0 \over 2\pi} {\cal A}(X,p) &=&
 -2 {\rm Im} \left(\oint\limits_{{\cal C}_1 + {\cal C}_2} 
\!\! {dp_0 \over 2\pi} \bar D^{\rm ret}(X,p)
- \int\limits_{{\cal C}_2 } \!\! {dp_0 \over 2\pi} \, {1 \over p_0}  \right)
= -2 {\rm Im} \left( 0 - {1 \over 2\pi} \int\limits_0^\pi \!\! d\phi \, i \right) = 1
\end{eqnarray}
where ${\cal C}_1$ is a path along the real axis and ${\cal C}_2$ is an infinitely
large half-circle surrounding counter clockwise the upper complex half-plane.

For the effective spectral function we note that $1-K$ can be written as 
(cf.~(\ref{eq:defK}))
\begin{equation}
  \label{eq:relrk}
1-K = \Gamma {\partial r \over \partial p_0}
\end{equation}
with the ratio
\begin{equation}
  \label{eq:defr2}
r(X,p) = {p_0 - {\vec p^2 \over 2m}-{\rm Re}\bar \Sigma^{\rm ret}(X,p)\over \Gamma(X,p)}
\,.
\end{equation}
Thus we get
\begin{equation}
  \label{eq:psspecr}
\tilde {\cal A} = {1 \over 2} \Gamma{\cal A}^2 (1-K) = 
{1 \over 2} {1 \over (r^2 + {1 \over 4})^2} {\partial r \over \partial p_0}   \,.
\end{equation}
Now the normalization of the effective spectral function can easily be shown:
\begin{equation}
  \label{eq:normpsspec}
\int\limits_{-\infty}^\infty \!\! {dp_0 \over 2\pi} \tilde {\cal A}
= \int\limits_{-\infty}^\infty \!\! {dr \over 2\pi} 
{1 \over 2} {1 \over (r^2 + {1 \over 4})^2} = 1  
\end{equation}
where we have used that
\begin{equation}
  \label{eq:limrpinf}
r  \to \pm \infty \qquad \mbox{for}  \qquad  p_0 \to \pm \infty  \,.
\end{equation}
Again we have assumed that for large $\vert p_0 \vert$ the self-energies can be neglected
as compared to $p_0$.


\section{Equation of motion for ${\cal S}$}  \label{app:eomstilde}

In this Appendix we will derive (\ref{eq:eomstilde}) from (\ref{eq:fineomsless}). 
We have to show that 
\begin{eqnarray}
\lefteqn{[p_0 - {\vec p^2 \over 2m}-{\rm Re}\bar \Sigma^{\rm ret}, \, {\cal S}]
- {1 \over \Gamma}  \,
[ \Gamma, \, (p_0 - {\vec p^2 \over 2m}-{\rm Re}\bar \Sigma^{\rm ret}){\cal S}] }
\nonumber \\
  \label{eq:diffvan}
&& {}-{1 \over 2} \Gamma {\cal A} \,
[p_0 - {\vec p^2 \over 2m}-{\rm Re}\bar \Sigma^{\rm ret}, \, S^<]
+ {1 \over 2} {\cal A} \,
[ \Gamma, \, (p_0 - {\vec p^2 \over 2m}-{\rm Re}\bar \Sigma^{\rm ret})S^<]
\end{eqnarray}
vanishes. Using the definition (\ref{eq:defstilde}) we find that (\ref{eq:diffvan})
reduces to
\begin{eqnarray}
\lefteqn{
[p_0 - {\vec p^2 \over 2m}-{\rm Re}\bar \Sigma^{\rm ret}, \, {1 \over 2} \Gamma {\cal A}]
\, S^<
- { p_0 - {\vec p^2 \over 2m}-{\rm Re}\bar \Sigma^{\rm ret} \over \Gamma}  \,
[ \Gamma, \, {1 \over 2} \Gamma {\cal A}] \, S^<  } \nonumber \\
&& = (p_0 - {\vec p^2 \over 2m}-{\rm Re}\bar \Sigma^{\rm ret}) \, S^< \,
[{\rm log}{p_0 - {\vec p^2 \over 2m}-{\rm Re}\bar \Sigma^{\rm ret} \over \Gamma} , \,
{1 \over 2} \Gamma {\cal A}]      \nonumber \\
  \label{eq:van2}
&& = (p_0 - {\vec p^2 \over 2m}-{\rm Re}\bar \Sigma^{\rm ret}) \, S^< \,
[{\rm log}(r) , \, {1 \over 2}{1 \over r^2 + {1 \over 4} } ]
\end{eqnarray}
with the ratio $r$ defined in (\ref{eq:defr2}).
To finish our proof we only have to note that the last Poisson bracket in 
(\ref{eq:van2}) vanishes on account of the following general property of the Poisson
bracket:
\begin{equation}
  \label{eq:proppois}
[f_1(g(X,p)), \, f_2(g(X,p))] = 0
\end{equation}
for arbitrary functions $g$ and arbitrary functions $f_1$, $f_2$ which do not explicitly
depend on $X$ and $p$ but only implicitly via $g$.


\begin{references}

\bibitem{Sc61} J.~Schwinger, J.~Math.~Phys. {\bf 2}, 407 (1961). 

\bibitem{KB} L.P.~Kadanoff and G.~Baym, `Quantum Statistical Mechanics',
Benjamin, New York (1962).

\bibitem{BM63} P.M.~Bakshi and K.T.~Mahanthappa, J.~Math.~Phys. {\bf 4}, 1, 12 
(1963). 

\bibitem{Ke64} L.V.~Keldysh, Zh.~Eks.~Teor.~Fiz. {\bf 47}, 1515 (1964);
Sov.~Phys.~JETP {\bf 20}, 1018 (1965).

\bibitem{Da84a} P.~Danielewicz, Ann.~Phys. {\bf 152}, 305 (1984).

\bibitem{Ch85} K.~Chou, Z.~Su, B.~Hao and L.~Yu, Phys.~Rep. {\bf 118},
1 (1985).

\bibitem{BotMal} W.~Botermans and R.~Malfliet, Phys.~Rep. {\bf 198}, 115 (1990).

\bibitem{DaM} S.~Mrowczynski and P.~Danielewicz, Nucl.~Phys. {\bf B342}, 345 (1990). 

\bibitem{MrHe} S.~Mrowczynski and U.~Heinz, Ann.~Phys. {\bf 229}, 1 (1994).

\bibitem{Mr} S.~Mrowczynski, Phys.~Rev.{\bf D56}, 2265 (1997).

\bibitem{GrLe98} C.~Greiner and S.~Leupold, Ann.~Phys. {\bf 270}, 328 (1998).

\bibitem{SL95} V.~\v Spi\v cka and P.~Lipavsk\'y, Phys.~Rev. {\bf B52}, 14615 (1995).

\bibitem{Kra} Th.~Bornath, D.~Kremp, W.D.~Kraeft, and M.~Schlanges, 
Phys.~Rev. {\bf E54}, 3274 (1996).

\bibitem{Reh} P.~Rehberg, Phys.~Rev. {\bf C57}, 3299 (1998).

\bibitem{henning} P.A.~Henning, Nucl.~Phys. {\bf A582}, 633 (1995), 
Erratum-ibid. {\bf A586}, 777 (1995); Phys.~Rep. {\bf 253}, 235 (1995).

\bibitem{IKV99} Yu.~Ivanov, J.~Knoll, and D.N.~Voskresensky, hep-ph/9807351,
nucl-th/9905028.

\bibitem{CJ99} W.~Cassing and S.~Juchem, nucl-th/9903070, 
to appear in Nucl.~Phys. {\bf A}.

\bibitem{EM99} M.~Effenberger, E.L.~Bratkovskaya, and U.~Mosel, Phys.~Rev. {\bf C60},
44614 (1999);
M.~Effenberger and U.~Mosel, Phys.~Rev. {\bf C60}, 51901 (1999).

\bibitem{kainu} K.~Kainulainen, private communication. 

\bibitem{kainu2} M.~Joyce, K.~Kainulainen, T.~Prokopec, hep-ph/9906413.

\bibitem{CJ992} W.~Cassing and S.~Juchem, nucl-th/9910052.

\end{references}
\end{document}